\documentclass[reprint,aps,prl,groupedaddress]{revtex4-2}

\usepackage{graphicx}
\usepackage{dcolumn}
\usepackage{amsmath}
\usepackage{bm}

\begin{document}


\title{Direct Observation of Chiral Phonons by Inelastic X-ray Scattering}


\author{Qingan Cai,\textsuperscript{1} Olle Hellman,\textsuperscript{3} Bin Wei,\textsuperscript{1,4} Qiyang Sun,\textsuperscript{1} Ayman H. Said,\textsuperscript{5} Thomas Gog,\textsuperscript{5} Barry Winn,\textsuperscript{6} }
\author{ Chen Li\textsuperscript{1,2}}
\email{chenli@ucr.edu}

\affiliation{\textsuperscript{1}Department of Mechanical Engineering, University of California at Riverside, Riverside, CA 92521 \\
\textsuperscript{2}Materials Science and Engineering, University of California at Riverside, Riverside, CA 92521\\
\textsuperscript{3}Department of Physics, Chemistry and Biology (IFM), Linköping University, SE-581 83 Linköping, Sweden\\
\textsuperscript{4}Henan Key Laboratory of Materials on Deep-Earth Engineering, School of Materials Science and Engineering, Henan Polytechnic University, Jiaozuo 454000, China\\
\textsuperscript{5}Advanced Photon Source, Argonne National Laboratory, Argonne, IL 60439\\
\textsuperscript{6}Neutron Scattering Directorate, Oak Ridge National Laboratory, Oak Ridge, TN 37831}


\date{\today}

\begin{abstract}
Phonon chirality has attracted intensive attention since it breaks the traditional cognition that phonons are linear propagating bosons. This new quasiparticle property has been extensively studied theoretically and experimentally. However, characterization of the phonon chirality throughout the full Brillouin zone is still not possible due to the lack of available experimental tools. In this work, phonon dispersion and chirality of tungsten carbide were investigated by millielectronvolt energy-resolution inelastic X-ray scattering. The atomistic calculation indicates that in-plane longitudinal and transverse acoustic phonons near K and K’ points are circularly polarized due to the broken inversion symmetry. Anomalous inelastic X-ray scattering by these circularly polarized phonons was observed and attributed to their chirality. Our results show that inelastic X-ray scattering can be utilized to characterize phonon chirality in materials and suggest that a revision to the phonons’ scattering function is necessary. 
\end{abstract}


\maketitle

Circularly polarized phonons, or called chiral phonons, exhibit eigenmodes with circular atomic vibrations and have been theoretically predicted and experimentally observed in some 2D materials~\cite{zhang2015chiral,zhu2018observation}. Phonon chirality plays a significant role in controlling the quantum state, generating thermal hall effect, and assisting intervalley or intravalley electron-phonon scatterings~\cite{chen2019entanglement,grissonnanche2020chiral,zeng2012valley}. Great efforts have been made to probe chiral phonons and related physical phenomena by infrared circular dichroism and Raman scattering~\cite{zhu2018observation,yin2021chiral,chen2015helicity,du2019lattice}. However, these methods remain indirect at best and do not allow characterizing chirality throughout the full Brillouin zone (BZ). To date, non-resonant meV-energy-resolution inelastic X-ray scattering (IXS) techniques have only been used to work with linear phonons. The dynamical structure factor of linear phonons is well predicted by the scattering function based on Born approximation~\cite{baron2009phonons,sakai2012soft,tornatzky2019phonon}. To our knowledge, no attempt has been made to measure chiral phonons by IXS.

\begin{figure*}
\includegraphics[width=\linewidth]{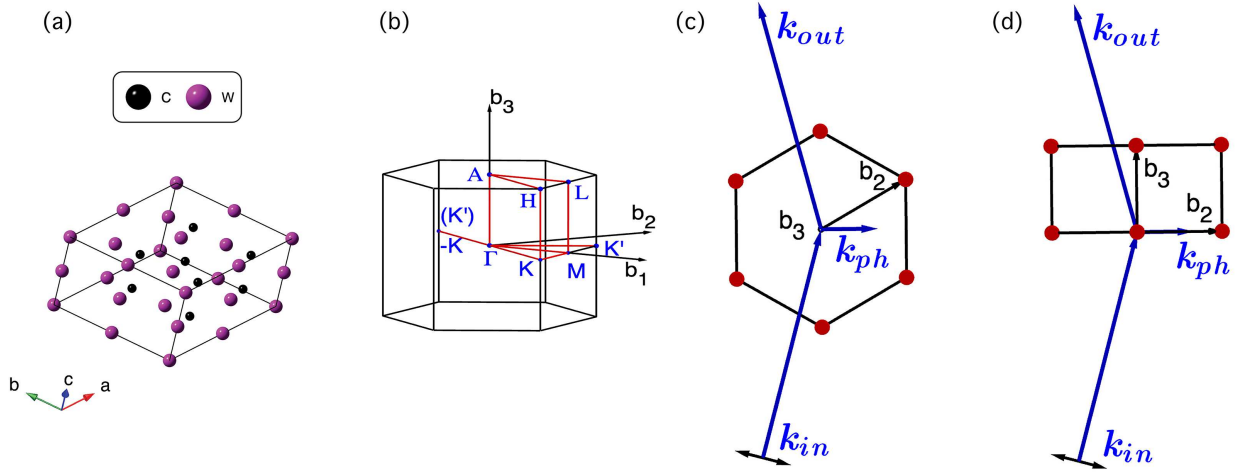}
\caption{\label{fig:wide1}(a) Tungsten carbide has hexagonal lattice structure. Purple and black spheres represent tungsten atoms and carbon atoms, respectively. (b) K and K$^\prime$ are distinguished in the Brillouin zone, indicating the broken inversion symmetry. (c)-(d) In-plane scattering geometry and out-of-plane scattering geometry are designed. Red balls in (c) and (d) represent the lattice points in reciprocal space; $k_{in}$, $k_{out}$ and, $k_{ph}$ represent the wavevector of incident electric filed, scattered electric filed and phonon during inelastic scattering process.}
\end{figure*}

To create phonon eigenstates with angular momentum, inversion symmetry needs to be broken~\cite{zhang2015chiral,coh2019classification}. Tungsten carbide (WC), with a space group of P$\bar{6}$m2 (No.187), has three-fold rotational symmetry with broken inversion symmetry as shown in FIG. \ref{fig:wide1} (a) and (b). The K point in BZ is not equivalent to its reversed K$^\prime$ point (-K) without inversion symmetry, as shown in Figure S1 (Supplemental Materials). Consequently, WC is an ideal candidate for investigating phonon chirality in 3-dimensional materials. Additionally, even though phonon transport properties of WC have been studied by first-principle calculations and phonon Boltzmann equation,~\cite{ma2018three,guo2018soft,kundu2020anomalously} no phonon measurement has been done to the best of our knowledge. Lattice dynamics measurements of WC would provide valuable information on its thermodynamic properties and phonon chirality.

In this letter, phonon measurement was performed by IXS on WC along several high symmetry directions in the BZ. The density of state (DOS) of acoustic phonons was also measured by inelastic neutron scattering (INS). Anomalous IXS by in-plane longitudinal (LA) and transverse acoustic (TA) phonons was observed at K and K$^\prime$ points. The anomaly was attributed to the chirality of these phonon modes. The discovery educes that phonon polarizations play significant roles in the X-ray scattering process. The results also provide insights on the utilization of IXS in probing phonon chirality and suggest that a revision of the scattering function to take into account chirality is necessary.

Phonon measurement was performed at the high energy-resolution inelastic X-ray (HERIX) spectrometer at 30-ID of the Advanced Photon Source, Argonne National Laboratory. Single crystals (typical size of ~200 $\times$ 300 $\mu$$m^2$) were purchased commercially (KENNAMETAL~\cite{kennametal}). The sample was attached to a copper post by GE varnish  and the copper post was mounted on a 4-axis rotation and 3-dimension translation stage. A photon energy of 23.7 keV (wavelength at 0.5226 Å) is used. The instrument has an energy resolution of 1.5 meV (full width at half maximum)~\cite{said2011new}. The beam is focused on a spot of 35 × 15 $\mu$$m^2$ (H $\times$ V). The thickness of the sample was optimized  for transmission measurements. The X-ray energy loss spectrum was collected by scanning the energy of high-resolution monochromator~\cite{toellner2011six}. CdTe Pilatus3 area detectors were used for data collection~\cite{said2020high}. All measurements were conducted at room temperature. In the scattering process, the incident X-ray beam is linearly polarized in the horizontal plane.

\begin{figure*}
\includegraphics[width=\linewidth]{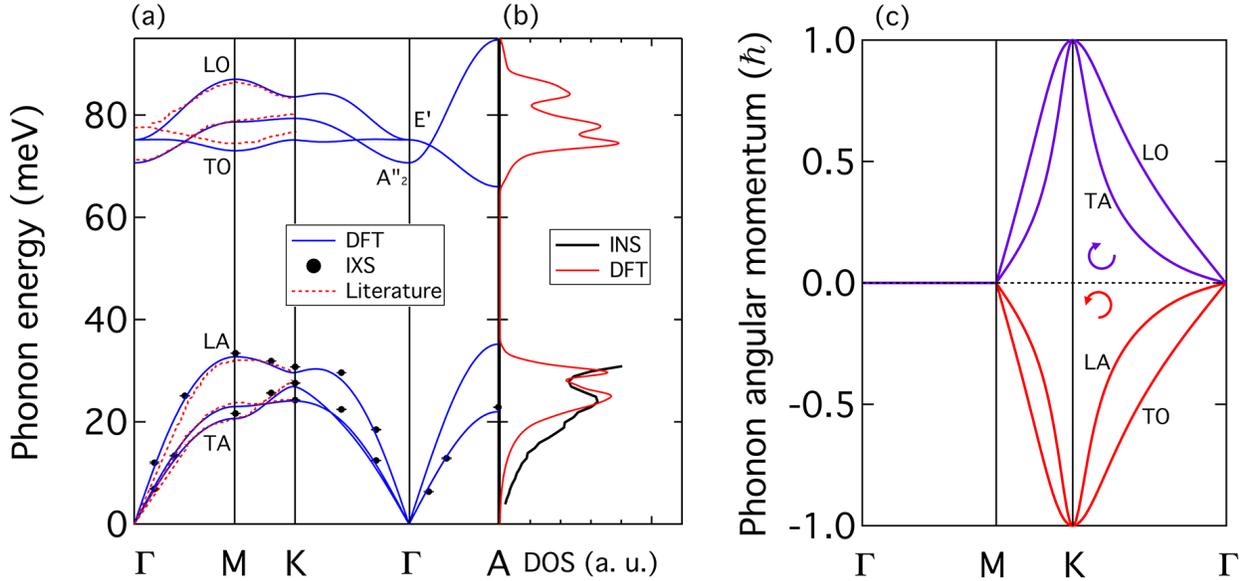}
\caption{\label{fig:wide2}(a) Phonon dispersion of WC from DFT calculation agrees well with experimental measurement. There are three acoustic and three optical phonons branches. Calculated phonon dispersion relations are plotted in lines and IXS measurement is represented by black symbols. The error bars indicate the fitting error of phonon energy in IXS spectra. The red dashed dispersion lines are from literature~\cite{guo2018soft}. (b)DFT calculation agrees well with measured acoustic phonon DOS of WC. The solid red line represents the calculated total DOS of WC; the black line represents the measured DOS below 30.9 meV.  (c) Phonons near K/K$^\prime$ points are circularly polarized. Phonon angular momentum is plotted by purple and red shadow on the phonon dispersion along high symmetry directions. The purple and red colors represent positive and negative value of angular momentum respectively.}
\end{figure*}

Phonon dispersion was measured along $\Gamma$–M–K–$\Gamma$–A directions in reciprocal space, shown in Figure S1 (b). FIG. \ref{fig:wide2} (a) shows the phonon dispersion of WC along these high symmetry directions. First-principles calculations are compared with these measurements with good agreement. First-principles calculations were performed with the density-functional theory (DFT) as implemented in Vienna Ab initio simulation package (VASP)~\cite{kresse1996efficient}. The exchange correlation function with the generalized gradient approximation in the Perdew-Burke-Ernzerhof flavor (GGA-PBE)~\cite{perdew1996generalized} and the projector-augmented-wave (PAW) potentials were used. Kinetic cutoff energy with 800 eV was used for plane wave expansion in reciprocal space with a k-point mesh of 12 × 12 × 12. The threshold for the total energy convergence was 10$^{-8}$ eV. The lattice constants obtained from relaxation (a = b = 2.911 Å, c = 2.859 Å) have a slightly larger value in c-axis compared to our experimental value (c = 2.844 Å, obtained from the Bragg peaks in IXS measurement). The phonon dispersion in harmonic approximation was calculated using Phonopy~\cite{togo2015first}. Second order force constants were calculated by the finite displacement method in supercells (3 × 3 × 3) containing 54 atoms. Gamma centered q-point meshes of 6 × 6 × 6 was used for each supercell.

The calculation agrees well with our IXS measurement of acoustic phonons, with some slight underestimation of the phonon energy along M–K–$\Gamma$ directions. Some energy difference can be found in transverse acoustic phonons at (2.2 0.2 0) in $\Gamma$–K direction and (1.8 0.6 0) in M–K direction. This underestimation may result from the slightly larger relaxed lattice parameters. Additionally, the current calculation agrees well with the experimental measurement along $\Gamma$–M direction, whereas the prior work from Ref.~\cite{guo2018soft} underestimates the phonon energy. Our calculation also produces lower optical phonon energy at $\Gamma$ point, especially for the double-degenerate E$^\prime$ mode at 75.16 meV (18.20 THz), which is 3\% less than Ref.~\cite{guo2018soft}. The extracted linewidths by super-resolution method are still considerably larger than expected from calculated phonon lifetime~\cite{guo2018soft} and indicate that IXS measurement is resolution limited for phonon linewidth of WC (Supplemental Materials). 

Acoustic part of the phonon DOS measured by inelastic neutron scattering (INS) also shows decent agreement with the calculation (FIG. \ref{fig:wide2}(b)). Powder INS measurement was performed on the time-of-flight direct geometry neutron spectrometer, Hybrid Spectrometers (HYSPEC), at the Spallation Neutron Source (SNS) at Oak Ridge National Laboratory. The measurement was performed at 300 K with incident energy ($E_i$) of 35 meV. The data were reduced using Mantid~\cite{arnold2014mantid}. The observed distinction comes from the instrument resolution and residual time independent background. The peak near 25 meV is dominated by TA branches and the peak about 30 meV corresponds to LA branch. In general, the IXS and INS measurements confirm the accuracy of the first-principles calculation for acoustic phonons.

According to the first-principles calculation, phonons at K and K$^\prime$ points have the largest angular momenta and are fully circularly polarized, as shown in FIG. \ref{fig:wide2}(c). Circular atomic vibrations of chiral phonons are shown by the time-dependent atomic displacement $u_l\left(t\right)$ is~\cite{sinha2001theory}:
\begin{eqnarray}
u_l\left(t\right)=\sum_{qj}&&{e_{qj}\left(\frac{\hbar}{2NM\omega_{qj}}\right)^\frac{1}{2}} \nonumber\\
&&\times 
{\left(\alpha_{qj}e^{-i\omega_{qjt}}+\alpha_{qj}^+e^{i\omega_{qjt}}\right)}.
\end{eqnarray}
where $\textit{l}$ goes through all atoms in the crystal, $\textit{N}$ is the number of atoms in the crystal, $\textit{M}$ is the atomic mass, and $\alpha_{qj}^+$ and $\alpha_{qj}$ are the phonon creation and annihilation operators, $e_{qj}$ is the eigenvector of jth phonon mode with reduced momentum at q, $\omega_{qj}$ is the corresponding eigenvalue. Three-dimensional eigenvectors of chiral phonons are complex. In-plane LA and TA modes are associated with the atomic motions in the basal plane, indicating that direction of angular momentum is along c-axis. Phonon angular momentum of mode j at q along c-axis, $l_{c}(q , j)$, can be calculated from Eq. 2~\cite{zhang2014angular,hamada2018phonon}: 
\begin{equation}
l_{c}(q , j)=\hbar(e_{qj}^\dag \mathbf{M}e_{qj}).
\end{equation}
Here \[\mathbf{M}=\left(\begin{matrix}0&-i&0\\i&0&0\\0&0&0\\\end{matrix}\right)\otimes I_{n\times n}\] and n is the number of atoms in one unit cell.

Near the K points (nonlinear region), TA and LO modes are dominantly clockwise circularly polarized, showing positive angular momentum along the c-axis; LA and TO modes are anticlockwise circularly polarized, showing negative angular momentum along c-axis. On the other hand, near the K$^\prime$ points, the phonon modes with the same frequency show opposite polarizations when compared with those near K points. The total angular momentum of these phonon modes throughout the full Brillouin zone is zero. TA and LA modes have opposite polarization at any given q point, so do LO and TO modes. 

When measuring chiral phonons, in-plane and out-of-plane scattering geometries were used to measure the scattering geometry-dependent dynamical structure factor of chiral phonons. IXS spectra at some K points ((0.33 2.33 0), (0.33 1.33 0) and (2.33 0.33 0))and K$^\prime$ point ((0.66 1.66 0)), as shown in Figure S1 (a), were obtained during the measurement in an in-plane scattering geometry, ( FIG. \ref{fig:wide1} (c)). Additionally, an out-of-plane scattering geometry in FIG. \ref{fig:wide1} (d) was used to measure the spectra at another two K$^\prime$ points ((2.66 -2.33 0) and (1.66 -2.33 0), shown in Figure S1 (a)). The same K$^\prime$ or K points are not possible in both in-plane and out-of-plane scattering geometries due to the limitations of rotational constrain and the points with equivalent $\left|Q\right|$ are used. For the in-plane scattering geometry, the basal plane (ab plane in real space shown in FIG. \ref{fig:wide1} (a)) of the crystal was oriented parallel to the plane of electric field vectors of the incident X-ray beam. For the out-of-plane scattering geometry, the basal plane of the crystal was oriented perpendicular to the plane of electric field vectors of the incident X-ray beam. 

Anomalous IXS by chiral phonons is observed at Brillouin zone boundary (K and K$^\prime$ points). Such anomaly depends on the scattering geometry. The results show that the scattering function based on the Born approximation reliably reproduces the scattering intensity of linearly polarized phonons and the circularly polarized phonons measured in the in-plane scattering geometry but fails in predicting the intensity of circularly polarized phonons measured in the out-of-plane scattering geometry. In order to qualify such anomaly,  phonon peaks in IXS spectra are fitted by Voigt functions and the area ratio between the LA and TA modes, $\frac{S_{LA}}{S_{TA}}$, is used to quantify the relative intensity between the two modes. The ratio is then compared with the one from the simulated dynamical structure factor, $S\left(Q,\omega\right)$, which is calculated from the partial differential cross section for the scattering process, as shown in Eq. 3~\cite{sinha2001theory}:
\begin{eqnarray}
\left(\frac{\partial^2\sigma}{\partial\Omega\partial E}\right)_{k\varepsilon_\alpha\rightarrow k^\prime\varepsilon_\beta}&&=\frac{k^\prime}{k}{r_e}^2\nonumber\\
&&\times
\left|{\varepsilon_\alpha}^\ast\cdot\varepsilon_\beta\right|^2S\left(Q,\omega\right).
\end{eqnarray}
where Q is the total wave vector transfer, $\omega$ is the phonon angular frequency. Photons are scattered from initial state with momentum k and polarization $\varepsilon_\alpha^\ast$  to a final state k$^\prime$ and $\varepsilon_\beta$, and $r_e$ is the electron radius. No change of photon polarization is involved and the partial differential cross section is directly proportional to the dynamical structure factor, S(Q,$\omega$). If considering one-phonon coherent scattering with photon energy loss (phonon creation) process, IXS intensities of LA and TA phonons were calculated by~\cite{baron2009phonons,sinha2001theory}:
\begin{eqnarray}
&&S\left(Q,\omega\right)_{1p}=N\sum_{q}\sum_{j}\sum_{d}\nonumber\\
&&\times
\left\{\left|\frac{f_d\left(Q\right)}{\sqrt{2M_d}}e^{-W_d\left(Q\right)}e^{iQ\cdot x_d}\left(Q\cdot e_{qjd}\right)\right|^2\right\}.
\end{eqnarray}
where N is the number of unit cells, q is the reduced momentum transfer determined by q=Q - $\tau$ ($\tau$ is the momentum transfer for Bragg reflections), d is the index of atoms in the primitive cell locate at $X_d$, j is the index of phonon modes, $M_d$ is the atomic mass, $f_d\left(Q\right)$ is the X-ray atomic form factor~\cite{brown2006intensity}, $\omega_{qj}$ is the phonon frequency, $e_{qjd}$ is the phonon polarization vector, $n_{\omega qj}$ is the Bose occupation, and the Debye Waller Factor is $e^{-2W_d\left(Q\right)}$. The shown simulated spectra are obtained from by using Eq. 4 are consistent with the results from Phonopy package. It should be noted that this formalism does not take the angular momentum of circularly polarized phonons into account.

\begin{figure}
\includegraphics[width=\linewidth]{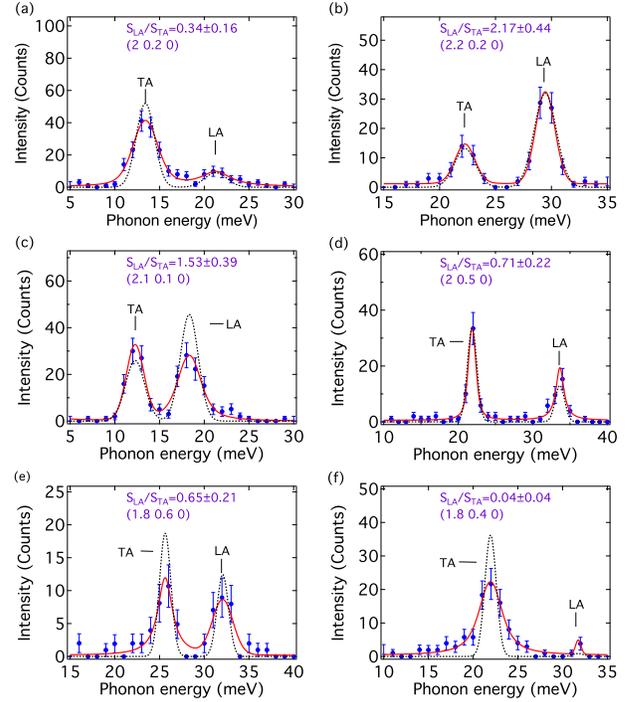}
\caption{\label{fig:wide3}The simulation based on Eq. 4 agrees well with the measurement at Q points away from K and K$^\prime$ points in the in-plane scattering geometry. (a)-(f) IXS spectra with linearly polarized X-ray at (2 0.2 0), (2.2 0.2 0), (2.1 0.1 0), (2 0.5 0), (1.8 0.6 0) and (1.8 0.4 0) in the reciprocal space, respectively. These spectra are obtained in the in-plane scattering geometry. Solid red lines are the Voigt fittings of IXS spectra and dotted black lines represent the spectra simulated from the scattering function. Error bars are from the counting statistics.}
\end{figure}

\begin{figure}
\includegraphics[width=\linewidth]{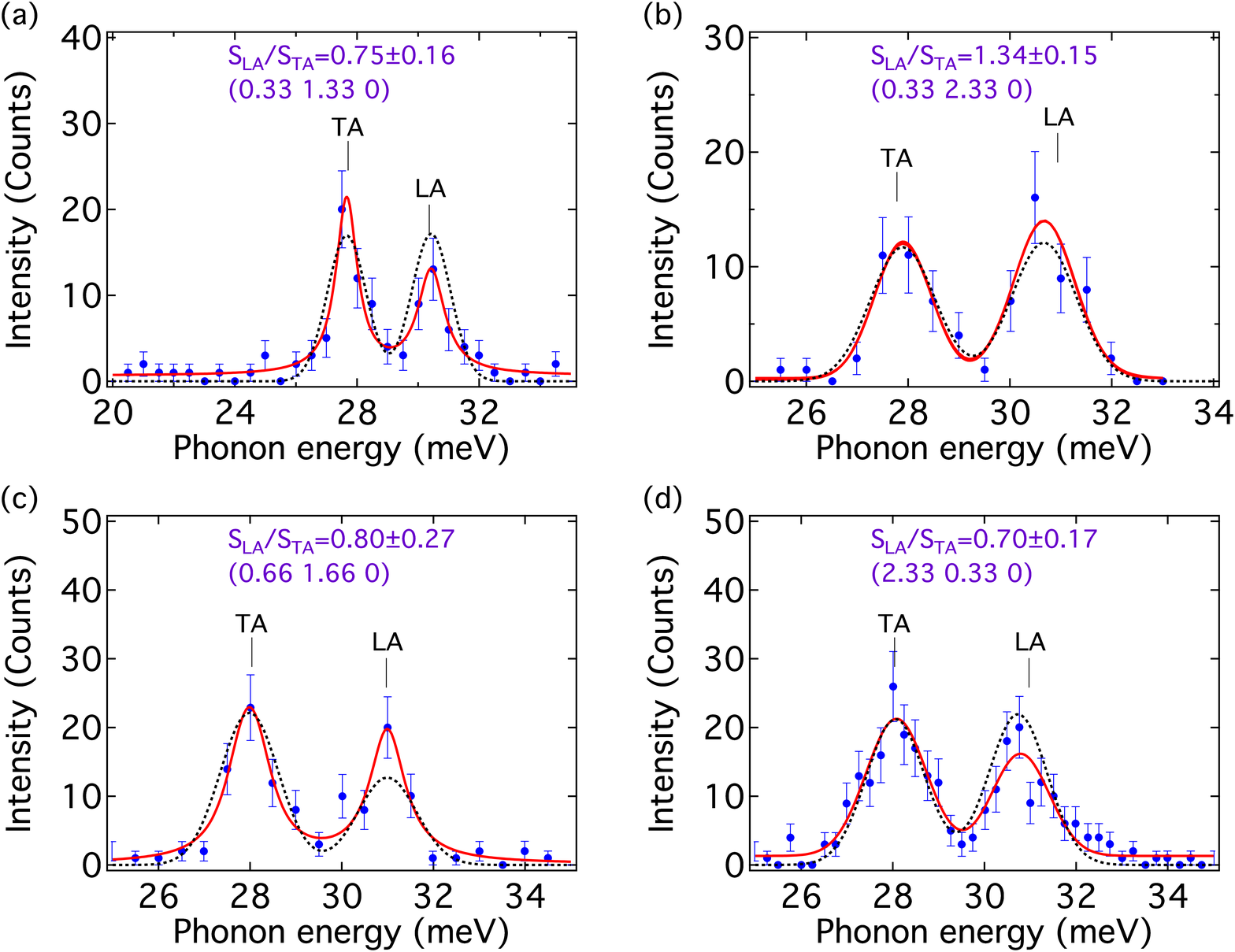}
\caption{\label{fig:wide4}(a) The simulation based on Eq. 4 agrees well with the measurement at K and K$^\prime$ points in the in-plane scattering geometry. (a)-(d) IXS spectra with linearly polarized X-ray at (0.33 1.33 0), (0.33 2.33 0), (0.66 1.66 0), and (2.33 0.33 0) in the reciprocal space, respectively. These spectra are obtained in the in-plane scattering geometry. Solid red lines are the Voigt fittings of IXS spectra and dotted black lines represent the spectra simulated from the scattering function. Error bars are from the counting statistics.}
\end{figure}

\begin{figure}
\includegraphics[width=\linewidth]{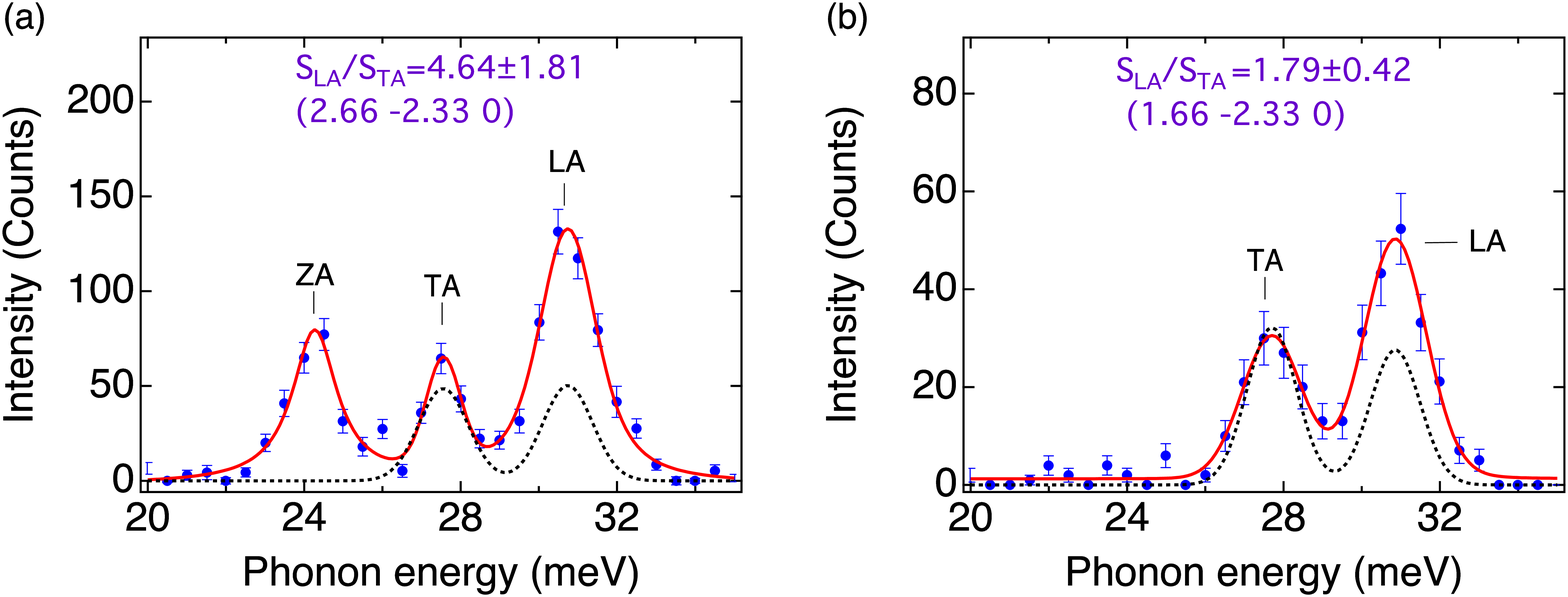}
\caption{\label{fig:wide5}The simulation based on Eq. 4 disagrees with the measurement at K and K$^\prime$ points in the out-of-plane scattering geometry. (a)-(b) IXS spectra at K$^\prime$ points (2.66 -2.33 0) and (1.66 -2.33 0); ZA is an out-of-plane acoustic phonon mode. These spectra are obtained in the out-of-plane scattering geometry. Solid red lines are the Voigt fittings of IXS spectra and dotted black lines represent the spectra simulated from the scattering function. Error bars are from the counting statistics.}
\end{figure}

At Q points away from K and K’, all simulation matches well with the measurement (FIG. \ref{fig:wide3}). At K and K$^\prime$ points in the in-plane scattering geometry, as shown in FIG. \ref{fig:wide4}, the simulation is consistent with the measurement as well. The similar intensities of TA and LA modes are generally predicted, within measurement statistics. However, for K$^\prime$ points in the out-of-plane scattering geometry in FIG. \ref{fig:wide5}, the measurement is significantly different from the simulation. It can be seen that intensity of LA mode at (2.66 -2.33 0) and (1.66 -2.33 0) is much stronger than that of  TA mode. 

For the same phonon mode, its scattering intensity is expected to be the same at (2.33 0.33 0), (0.33 2.33 0), and (2.66 -2.33 0) because they have equivalent $\left|Q\right|$ transfer and their phonon eigenvectors are in the same magnitude. Consequently, the intensity relation between LA and TA should be consistent in the three specified points. It should be noticed that the two phonon modes are fully circularly polarized at these K and K$^\prime$ points (FIG. \ref{fig:wide2}(b)). The obvious discrepancy between simulation and experiment indicates that the phonon chirality can be characterized by IXS. The discrepancy also reveals that Eq. 4 fails in reproducing the intensities of chiral phonons in the out-of-plane scattering geometry.

The failure of the scattering simulation shown here possibly originates from two sources. Firstly, the calculated cross section, Eq. 3, assumes that photon polarizations $\varepsilon_\alpha$ and $\varepsilon_\beta$  are conserved before and after the scattering by phonons, which works well for linear phonons. However, when phonons carry angular momentum, the creation or annihilation process may involve the transfer of angular momentum between photons and phonons. In this case, the photon polarizations $\varepsilon_\alpha$ and $\varepsilon_\beta$ might not be the same. Secondly, in Eq. 4, $S\left(Q,\omega\right)$ is proportional to $\left|(Q\cdot e_{qjd})\right|^2$, in which the eigenvector of the phonon mode with angular momentum is complex, containing phase information. Such information is lost in the calculation. Therefor, the eigenvectors of the similar magnitude leads to similar LA and TA intensities at K and K$^\prime$ points indifferent of the scattering geometry. 

In reality, because this IXS anomaly depends on scattering geometry. It is likely the phonon angular momentum is involved in scattering of photons. In the in-plane scattering geometry, the eigenvectors of LA and TA phonons are parallel to the electric field vectors of incident photons. In this scattering geometry, the phonon angular momentum $l_{qj}$ is perpendicular to the electric field vector of the photon. This scattering geometry may forbid the direct angular momentum transfer between phonons and photons. In the out-of-plane scattering geometry, the two directions are not perpendicular and angular momentum transfer between phonons and photons may be allowed. The result hints that the IXS anomaly could arise from the chiral phonon and polarization of phonon/X-ray may play an essential role in the scattering process. X-ray beam may carry angular momentum as spin angular momentum (SAM) and orbital angular momentum (OAM). It may be speculated that there are exchanges of SAM and/or OAM between phonons and photons. 


The anomalous scattering of linearly polarized X-ray by chiral phonons may involve the SAM and/or OAM between phonons and X-ray/photons. Linearly polarized X-ray takes a hybrid state where left- and right-hand polarized components are in phase. In IXS process, X-ray may be scattered differently by oppositely polarized TA and LA phonons, leading to their intensity difference. In addition, such interaction may be attributed to the current-current correlation function. Some major questions remain unanswered. The scattering function Eq. 4 needs to be revised to explain the IXS anomaly. The transfer of angular momentum between lattice and photons, if possible, should be considered. The answers to these questions will enable the inelastic X-ray scattering as a tool for quantify chiral phonons and their interactions with other phonons and other degrees of freedom for applications in valleytronics and quantum devices.

In summary, the acoustic phonons of WC along high symmetry directions have been measured by inelastic X-ray scattering and acoustic phonon DOS has been measured by inelastic neutron scattering. DFT calculations are compared these measurements with good agreement. The phonon angular momentum calculation shows that phonons are fully circularly polarized at zone boundary. While the atomistic simulation works well for most LA and TA phonons elsewhere, it fails dramatically for chiral phonons in out-of-plane geometry. It suggests that the anomalous intensity difference between these two phonon modes may arise from their opposite angular momentum. The observed IXS anomaly from chiral phonons suggests that the revision of scattering function for phonons is necessary and lays a foundation of future use of IXS as a tool for investigations on chiral phonons and the related interactions.  

C.L. and Q.C. acknowledge the support of University of California, Riverside via Initial complement. This research used resources of the Advanced Photon Source, a U.S. Department of Energy (DOE) Office of Science User Facility, operated for the DOE Office of Science by Argonne National Laboratory under Contract No. DE-AC02-06CH11357. Research at Oak Ridge National Laboratory’s SNS was sponsored by the Scientific User Facilities Division, BES, DOE.

\nocite{*}
\bibliographystyle{apsrev4-2}
\bibliography{reference}
\end{document}